# An Analytical Solution of the Equations of a Rolling Disk of Finite Tickness on a Rough Plane

Milan Batista

University of Ljubljana, Faculty of Maritime Studies and Transportation

Pot pomorscakov 4, 6320 Portoroz, Slovenia, EU

milan.batista@fpp.edu

**Abstract**

In this article an analytical solution of equations of motion of a rigid disk of finite thickness rolling on its edge on a perfectly rough horizontal plane under the action of gravity is given. The solution is given in terms of Gauss hypergeometrical functions.

## 1  Introduction

This paper is devoted to an integration of the equations of motion of a rigid disk of finite thickness rolling on its edge on a perfectly rough horizontal plane under the action of gravity. Historically, the problem was one of the challenges of 19 century dynamics. From the historical notes of Routh ([9]), O'Reilly ([8]) and Borisov et al ([3]), the solution of the problem for an infinitely thin disk in the terms of Gauss hypergeometric functions was in 1900 provided by P.Appel and D.Korteweg and before them independently, in 1897, by S.A.Chaplygin. Their solution can be found, for example, in textbooks (see [1],[7],[11]). A little later, in 1903, E.Gallop noted that the solution of the problem of a rolling disk leads to Legendere's equation. His solution can be found in Routh's book ([9]).  Recently, the analytical solution in terms of Legendre functions was used by O'Reilly ([8]) for the study of bifurcations and stability of steady motion of an infinitely thin disk. A similar study was performed by Kuleshev ([5],[6]) but he used solutions in terms of  Gauss hypergeometric functions. These functions were also used by Borisov at all ([3]) for construction of the bifurcation diagrams and qualitative analysis of the point of contact of a disk on a plane.

From the review of literature it is clear that the motion of an infinitely thin disk is well studied, but this is not the case for disk of finite thickness. As was noted by Routh ([9]), the equation of motion for that case can be derived from the general equation of motion of the body of revolution by putting the radius of curvature at zero. This was done by Webster ([11]), but he then treated only the special case of an infinitely thin disk. It seems that only the recent paper of Kessler and O'Reilly ([4]), devoted to the settling of a science toy called Euler's disk, discusses a disk of finite thickness. In this paper the authors derived the equation of motion for the case of sliding and that of rolling and rolling with slip and then performed a numerical simulation of the so-called stick-slip movement of the disk.

The purpose of this paper is to provide an analytical solution of the equation of motion of the problem which as it seems has not yet been given. The next section reviews the basic equations, the following gives the solution of the equations and the last provides an example of the application of the derived solution through the calculation of phase diagrams and normal reaction force.

## 2 Equations

Consider a rigid homogeneous disk moving on a perfectly rough horizontal plane in the homogeneous gravity field with acceleration $g$. The disk radius is $a$, height $2h$ and mass $m$. In what follows, the dimensionless form of the equations will be used. They are obtained if units of mass, length and time are $m$, $a$, and $\sqrt{a/g}$.

Following Kessler and O'Reilly ([2],[4]), the position of the disk is defined by the Cartesian coordinates of its centre of mass and the Euler angles $(\psi,\theta,\varphi)$ where $\psi$ is the precession angle, $\theta \in (-\pi/2, \pi/2,)$ is the inclination angle--i.e., the angle between the disk axis and the plane and vertical--and $\varphi$ is the rotation angle. Note that the present notation differs from that of [4] where Euler's angles are denoted by $(\theta,\alpha,\psi)$. If

$$\frac{d\psi}{dt}=\frac{\omega_3}{\cos\theta} \qquad \frac{d\theta}{dt}=\omega_1 \qquad \frac{d\varphi}{dt}=\omega_2-\omega_3\tan\theta \tag{1}$$

then the equation of the motion of a disk can be reduced to the following system of equations

$$\begin{aligned}
\frac{d\omega_1}{dt} &= \frac{1}{(15+16h^2)}\left\{6\left(3+2\tilde{h}\tan\theta\right)\omega_2\omega_3 - \left[\left(3+16h^2\right)\tan\theta+12\tilde{h}\right]\omega_3^2 \right. \\
&\quad \left. +12\left(\sin\theta-\tilde{h}\cos\theta\right)\right\} \\
\frac{d\omega_2}{dt} &= -\frac{1}{(3+8h^2)}\left[4\tilde{h}\omega_2+\frac{2}{3}\left(3+4h^2\right)\omega_3\right]\omega_1 \\
\frac{d\omega_3}{dt} &= \left[-\frac{6}{3+8h^2}\omega_2+\left(\tan\theta+\frac{4\tilde{h}}{3+8h^2}\right)\omega_3\right]\omega_1
\end{aligned} \tag{2}$$

where $\tilde{h}\equiv h\,\mathrm{sgn}(\theta)$ and $\theta\in(-\pi/2,\pi/2,)/\{0\}$. The first integral of (2) is the energy integral which is given by

$$E=\left(\frac{5}{8}+\frac{2}{3}h^2\right)\omega_1^2+\frac{3}{4}\omega_2^2-\tilde{h}\omega_2\omega_3+\left(\frac{1}{8}+\frac{2}{3}h^2\right)\omega_3^2+\cos\theta+\tilde{h}\sin\theta \tag{3}$$

Once $\omega_2$ and $\omega_3$ are known, $\omega_1$ is determinate from (3). To obtain equations for $\omega_2$ and $\omega_3$ the variables are changing from $t$ to $\theta$. By noting that $\frac{d}{dt}=\frac{d}{d\theta}\frac{d\theta}{dt}=\omega_1\frac{d}{d\theta}$ the equations $(2)_{2,3}$ become

$$\begin{aligned}
\left(3+8h^2\right)\frac{d\omega_2}{d\theta}+4\tilde{h}\omega_2+\frac{2}{3}\left(3+4h^2\right)\omega_3 &= 0 \\
\left(3+8h^2\right)\frac{d\omega_3}{d\theta}+6\omega_2-\left[2\left(3+8h^2\right)\tan\theta+4\tilde{h}\right]\omega_3 &= 0
\end{aligned} \tag{4}$$

These equations form a system of two homogeneous linear differential equations for unknown $\omega_2$ and $\omega_3$. From the first, one can express

$$\omega_3 = -\frac{6\tilde{h}}{(3+4h^2)}\omega_2 - \frac{3(3+8h^2)}{2(3+4h^2)}\frac{d\omega_2}{d\theta} \qquad (5)$$

and substituting this into the second, yield linear a second order differential equation for unknown $\omega_2$

$$\frac{d^2\omega_2}{d\theta^2} - \tan\theta\frac{d\omega_2}{d\theta} - K(1+\tilde{h}\tan\theta)\omega_2 = 0 \qquad (6)$$

where $K \equiv \dfrac{4}{3+8h^2}$. When (6) is solved, $\omega_3$ can be obtained from (5) and $\omega_1$ from the energy integral (3).

## 2 Solution

Equation (6) will now in two steps be reduced to the Gauss hypergeometric differential equation. First, if the new variable is introducing

$$s = \tan\theta \qquad (7)$$

then $\dfrac{d\omega_2}{d\theta} = \dfrac{d\omega_2}{ds}\dfrac{ds}{d\theta} = (1+s^2)\dfrac{d\omega_2}{ds}$, so (6) is transformed to

$$(1+s^2)^2\frac{d^2\omega_2}{ds^2} + s(1+s^2)\frac{d\omega_2}{ds} - K(1+\tilde{h}s)\omega_2 = 0 \qquad (8)$$

Equation (8) is a special case of the Riemann-Papperitz ([10]) equation with two singular points. The function

$$\sigma = \frac{s-\hat{i}}{s+\hat{i}} \tag{9}$$

where $\hat{i} = \sqrt{-1}$, maps singular points $s = -\hat{i}, \hat{i}$ to points $\sigma = 0, \infty$. By using (9) and the chain rule $\frac{d\omega_2}{ds} = \frac{d\omega_2}{d\sigma}\frac{d\sigma}{ds} = \frac{2\hat{i}}{(1-\sigma)^2}\frac{d\omega_2}{d\sigma}$ equation (8) is transformed to

$$4\sigma^2(\sigma-1)\frac{d^2\omega_2}{d\sigma^2} + 2\sigma(3\sigma-1)\frac{d\omega_2}{d\sigma} + K(\bar{\alpha}\sigma - \alpha)\omega_2 = 0 \tag{10}$$

where $\alpha \equiv 1+\hat{i}\tilde{h}$ and $\bar{\alpha} \equiv 1-\hat{i}\tilde{h}$. Here in what follows a bar over variable denotes its conjugate complex value. The solution of (10) is assumed to be of the form ([10])

$$\omega_2 = \sigma^\lambda Y(\sigma) \tag{11}$$

where $\lambda$ is constant and $Y(s)$ is a new function both of which must be determinate. Substituting (11) into (10) and setting

$$\lambda^2 - \frac{\lambda}{2} + \frac{K\alpha}{4} = 0 \tag{12}$$

Yields the hypergeometric equation

$$\sigma(1-\sigma)\frac{d^2Y}{d\sigma^2} + \left[\left(\frac{1}{2}+2\lambda\right)-\left(\frac{3}{2}+2\lambda\right)\sigma\right]\frac{dY}{d\sigma} - \left(\lambda^2 + \frac{\lambda}{2} + \frac{K\bar{\alpha}}{4}\right)Y = 0 \tag{13}$$

This is the Gauss hypergeometric equation which has the solution ([10], [12])

$$Y = C_1 F(a,b,c,\sigma) + C_2 \sigma^{1-c} F(a-c+1, b-c+1, 2-c, \sigma) \tag{14}$$

where $F$ is the Gauss hypergeometric function, $C_1, C_2$ are arbitrary complex constants and

$$c = a + b = \frac{1}{2} + 2\lambda \qquad ab = \lambda^2 + \frac{\lambda}{2} + \frac{K\bar{\alpha}}{4} \qquad (15)$$

As seen from (15) $a$ and $b$ are the roots of the quadratic equation $x^2 - (a+b)x + ab = 0$. By selecting the root of (12) to be

$$\lambda = \frac{1+\Delta}{4} \qquad \Delta \equiv \sqrt{1-4K\alpha} = \mu + \hat{i}\nu \qquad (16)$$

$a$ and $b$ are given by

$$a = \frac{1+\mu}{2} \qquad b = \frac{1+\hat{i}\nu}{2} \qquad (17)$$

Note that in this case $a$ is a real parameter. The solution of (8) can now, by using (11), (16) and (17) and noting that, by (7), $\frac{s-\hat{i}}{s+\hat{i}} = -e^{2\hat{i}\theta}$ be written in the compact form

$$\omega_2 = C_1 T_{\mu,\nu}(\theta) + C_2 T_{-\mu,-\nu}(\theta) \qquad (18)$$

where $T_{\mu,\nu}(\theta)$ is a two parameter complex valued function defined as

$$T_{\mu,\nu}(\theta) \equiv (-1)^{\frac{1+\mu+\hat{i}\nu}{4}} \left(e^{2\hat{i}\theta}\right)^{\frac{1+\mu+\hat{i}\nu}{4}} F\left(\frac{1+\mu}{2}, \frac{1+\hat{i}\nu}{2}, 1+\frac{\mu+\hat{i}\nu}{2}, -e^{2\hat{i}\theta}\right) \qquad (19)$$

with its first derivative given by

$$\frac{dT_{\mu,\nu}}{d\theta} = \hat{i}\frac{1+\mu+\hat{i}\nu}{2}T_{\mu,\nu}(\theta) + \hat{i}\left(-e^{2\hat{i}\theta}\right)\frac{(1+\mu)(1+\hat{i}\nu)}{(2+\mu+\hat{i}\nu)}T_{\mu+1,\nu+1}(\theta) \qquad (20)$$

By (18) the problem is formally solved and can be used for actual calculations. However, since only the real part of the solution is necessary, (18) will be further transformed. For the real solution one must have

$$\begin{aligned}\omega_2 &= C_1 T_{\mu,\nu} + C_2 T_{-\mu,-\nu} = \overline{C}_1 \overline{T}_{\mu,\nu} + \overline{C}_2 \overline{T}_{-\mu,-\nu} \\ \omega_2' &= C_1 T'_{\mu,\nu} + C_2 T'_{-\mu,-\nu} = \overline{C}_1 \overline{T}'_{\mu,\nu} + \overline{C}_2 \overline{T}'_{-\mu,-\nu}\end{aligned} \quad (21)$$

where $(\ )' = \dfrac{d}{d\theta}$. Eliminating $C_2$ from those equations and defining the new two parameters complex valued function

$$Z_{\mu,\nu}(\theta) \equiv \frac{\left(T_{-\mu,-\nu} T'_{\mu,\nu} - T_{\mu,\nu} T'_{-\mu,-\nu}\right)\overline{T}_{-\mu,-\nu}}{T_{-\mu,-\nu}\overline{T}'_{-\mu,-\nu} - \overline{T}_{-\mu,-\nu} T'_{-\mu,-\nu}} \quad (22)$$

one finds that $\omega_2 = C_1 Z_{\mu,\nu} + \overline{C}_1 \overline{Z}_{\mu,\nu}$. By putting $C_1 = \dfrac{A - \hat{i}B}{2}$ where $A$ and $B$ are real arbitrary constants, this can be rewritten into the form

$$\omega_2 = A M_{\mu,\nu} + B N_{\mu,\nu} \quad (23)$$

where $M_{\mu,\nu}$ and $N_{\mu,\nu}$ are real and imaginary parts of $Z_{\mu,\nu}$ given by

$$M_{\mu,\nu} \equiv \frac{Z_{\mu,\nu} + \overline{Z}_{\mu,\nu}}{2} \quad \text{and} \quad N_{\mu,\nu} \equiv \frac{Z_{\mu,\nu} - \overline{Z}_{\mu,\nu}}{2\hat{i}} \quad (24)$$

The solution (23) is the final solution of the equation (6) but not of the problem of motion of the disk because the case $\theta = 0$ is not covered and the presence of the function $\tilde{h} \equiv h\,\text{sgn}(\theta)$, which, by (16), determines only the sign of $\nu$. The final solution of the problem for $\theta \in (-\pi/2, \pi/2,)/\{0\}$ is therefore given by

$$\omega_2 = \begin{cases} A^+ M_{\mu,\nu} + B^+ N_{\mu,\nu} & 0 < \theta < \pi/2 \\ A^- M_{\mu,-\nu} + B^- N_{\mu,-\nu} & -\pi/2 < \theta < 0 \end{cases} \quad (25)$$

The connection between constants $A^+, B^+$ and $A^-, B^-$ depends on the condition of imposing on $\omega_2$ and it's derivative $\theta = 0$.

### 3 Solution at $\theta = 0$

Physically, when passing $\theta = 0$, the disk impacts the plane on a line on the disk bounding surface. However if one assumes that the bounding surface is slightly concave, then at $\theta = 0$ the disk contacts the plane on two points on each edge of each side. Thus at impact the disk loses contact at one edge and gains it on the other. Omitting the details it can easily be shown from the general equation of motion of the disk ([2]) that in this case the components of angular velocities after the impact are

$$\omega_1^+ = \frac{15 - 8(1+3\varepsilon)h^2}{15 + 16h^2}\omega_1^- \qquad \omega_2^+ = \omega_2^- - \frac{4\tilde{h}(3+4h^2)}{3(3+8h^2)}\omega_3^- \qquad \omega_3^+ = \frac{3}{3+8h^2}\omega_3^- \quad (26)$$

where $+$ and $-$ superscripts denote values after and before impact and $\varepsilon$ is the restitution coefficient. Using (26) it follows from (3) that the energy lost at the impact is

$$\Delta E = -2h^2\left[(1-\varepsilon^2)\frac{15+4h^2}{15+16h^2}\omega_1^{-2} + \frac{3+4h^2}{3(3+8h^2)}\omega_3^{-2}\right] \quad (27)$$

It is clear from (27) that when passing $\theta = 0$ there is always energy loss even in the case of elastic impact; i.e., when $\varepsilon = 1$ providing $h > 0$ and $\omega_3^- \neq 0$. In the case of an infinitely thin disk when $h = 0$ it follows from (26) that the component of angular velocities pass continuously through $\theta = 0$ and also, from (27), that there is no loss of energy; i.e., $\Delta E = 0$.

## 4 Example

Once the analytical solution is known it can be used for various purposes; for example, for studies of steady motion and bifurcations. In this paper, however, it will be used to illustrate the calculation of the phase diagram and calculation of the normal force which is given by ([2],[3])

$$F_z = 1 - \left(\cos\theta + \tilde{h}\sin\theta\right)\omega_1^2 - \left(\sin\theta - \tilde{h}\cos\theta\right)\frac{d\omega_1}{dt} \qquad (28)$$

Note that unilateral constraint requires that $F_z > 0$.

For the purpose of numerical calculation the functions (24) and their derivatives were tabulated for various values of $h$ and $\theta$ by using the Maple program. The tabulated values were then used in a special written Fortran program was used for various calculations. Here it is worthwhile to note that there is practically no program in the public domain for an evaluation of Gauss hypergeometric functions with complex parameters and argument. (In fact, the author found just two; one of which provided incorrect results, while the other was extremely slow).

Figures 1 and 2 show the phase portraits of (3) and contour graph of normal force given by (28) for the two case of $h = 0.17$. For both cases the integration constants $A$ and $B$ were calculated from the condition of steady motion with $\theta_0 = \arctan h$ and giving spin angular velocity, which is defined as $\Omega \equiv d\psi/dt$ ([2]).

In the case $\Omega_0 = 0.1$ shown in Figure 1 the calculated values of the integration constant are $A^+ = -1.175 \times 10^{-2}$ and $B^+ = -2.571 \times 10^{-2}$ (approximately). For $\theta < 0$ the constants are, from (26), $A^- = -0.166$ and $B^- = -0.134$. It is seen from Figure 1 (left) that on each side of the $\theta = 0$ line there are one stable and one unstable state. When passing the line $\theta = 0$ energy drops (cf. (27)) and the resulting bouncing settles the disk at the stable position $\theta = 0$. The contour plot of normal force shows that the whole theory is valid only for $|\omega_1|<1$, approximately, since for larger values of $\omega_1$ the normal force becomes negative.

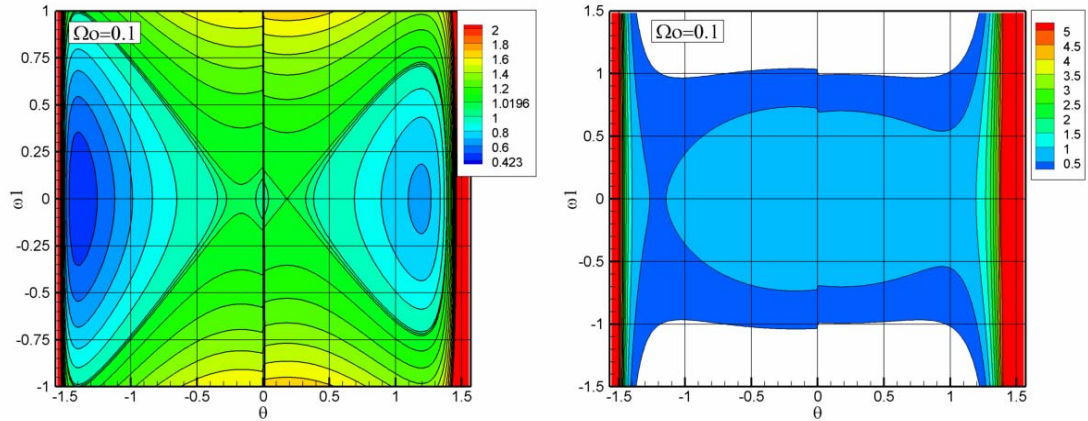

**Figure 1.** Phase portraits of (3) (left) and contour plot of normal force (right) for $h=0.17$ and $\Omega_0 = 0.1$.

In the case $\Omega_0 = 1.0$ shown in Figure 2 the calculated values of the integration constant are $A^+ = -6.953\times 10^{-2}$ and $B^+ = -0.134$ and, from (26), $A^- = 2.830$ and $B^- = 2.816$ (approximately). Unlike the previous case as it is seen from Figure 2 (left) there is no stable position on the negative side of $\theta = 0$ and only one stable position on the positive side. In each passing of the line $\theta = 0$ the energy drops (cf. (27)) and the resulting bouncing settles the disk in the orbit just touching $\theta = 0^+$ around the stable position. Again the contour plot of normal force [Figure 2 (right)] shows that the whole theory is valid only for $|\omega_1|<1$, approximately, since for larger values of $\omega_1$ the normal force becomes negative.

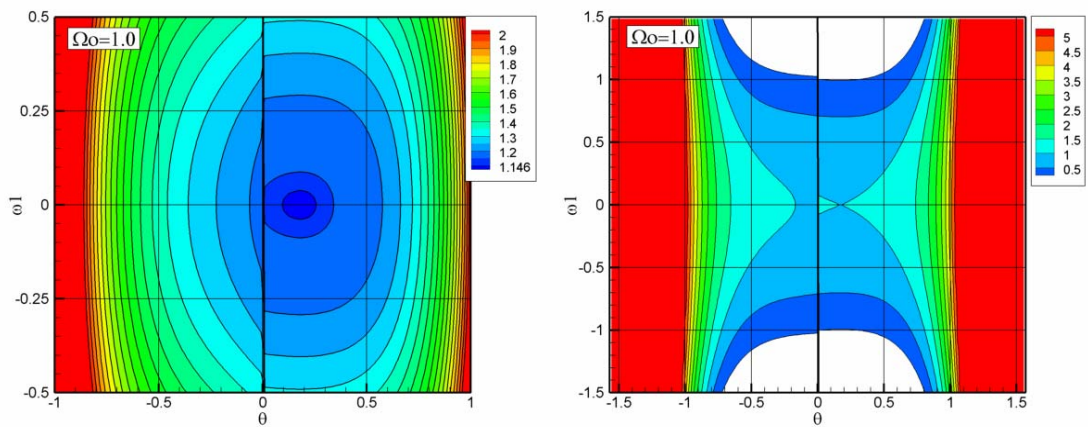

**Figure 2.** Phase portraits of (3) (left) and contour plot of normal force (right) for $h = 0.17$ and $\Omega_0 = 1$

## Conclusions

An analytical solution of equations of motion of a rigid disk of finite thickness rolling on its edge on a perfectly rough horizontal plane under the action of gravity is given in terms of Gauss hypergeometrical functions. Unlike with an infinitely thin disk, the solution has its jump at point $\theta = 0$ due to the impact of disk with the plane. Examples worked out show that the solution correctly predicts the vertical stable equilibrium position for a slowly rotated disk. It also shows that the normal force can become zero for $|\omega_1| > 1$, approximately, so the disk could lose contact with surface. This is, however, not the case for larger inclination angles where normal contact force grows rapidly.